# A Philosophical Argument Against Time Machines[1]

## Juliano C. S. Neves

**Introduction**

    General relativity is the theory of space, time, and the gravitational phenomenon, which is generated by both matter and energy. The most important Albert Einstein's work has been tested and confirmed so far. The most recent test was the detection of gravitational waves by the international collaboration LIGO (The Laser Interferometer Gravitational-Wave Observatory).[2] During the 20th and 21th centuries, general relativity has obtained successful results and reliability. But should every prediction in general relativity be considered reliable? Alongside general relativity's predictions such as gravitational waves, black holes, and the hypothetical big bang,[3] the Einsteinian theory provides special space-time curves

---

[1] Time machines are predictions of Einstein's theory of general relativity and provide a myriad of unsolved paradoxes. Convincing and general arguments against time machines and their paradoxes are missing in physics and philosophy so far. In this article, a philosophical argument against time machines is given. When thought of as a process, individuation refuses the idea of time machines, in particular travels into the past. With the aid of Nietzsche-Heraclitus' philosophy of becoming and Simondon's notion of process of individuation, I propose that time machines are modern fables, created by the man of *ressentiment*. In the *amor fati* formula of Nietzsche, I suggest the antipode to time machines.

This study was financed in part by the Coordenação de Aperfeiçoamento de Pessoal de Nível Superior—Brazil (CAPES)—Finance Code 001.

[2] Abbott, B. P. et al. (LIGO Scientific Collaboration and Virgo Collaboration). "Observation of gravitational waves from a binary black hole merger." *Physical Review Letters* 116: 061102, 2016.

[3] The big bang is not the unique response regarding the origin of the universe because bouncing cosmologies—models without the big bang—are possible in science today. See, for example, Brandenberger, R., and Peter, P. "Bouncing cosmologies: progress and problems." *Foundations of Physics, 47* (6): 797-850, 2017, Novello, M. and Perez Bergliaffa S. E. "Bouncing cosmologies." *Physics Reports* 463 (4): 127-213, 2008, and Neves, J. "O eterno retorno hoje." *Cadernos Nietzsche* 32: 283-296, 2013, and "Bouncing cosmology inspired by regular black holes." *General Relativity and Gravitation* 49 (9): 124, 2017.

called closed time-like curves (CTCs). Time-like curves are natural paths of observers, i.e., humans and every object with mass travel in space-time through time-like curves. And, in particular, closed time-like curves are curves or paths where, according to general relativity, observers would travel into the past or into the future. Then, the question in general relativity is not merely time dilation or different elapsed times generated by relative motion of observers as described by special relativity. The main question in general relativity is geometric, that is to say, different space-times (also known as geometries in Einstein's theory of gravitation) may provide CTCs and—at least mathematically—a direct form to travel in time. Then, among researches in general relativity, CTCs mean "time machines."

Time machines are amongst the most interesting and attractive subjects (especially for the general public) in theoretical physics. In general relativity, the possibility of returning into the past brings paradoxes like the problem, for an observer, of traveling into the past to kill, for example, his own grandfather. There exist some physical and philosophical arguments that try to avoid such paradoxes like the grandfather paradox. For example, Hawking [4] with his chronology protection conjecture tries to avoid the paradoxes and causality violations that involve time machines. Hawking uses, among his arguments in order to reject CTCs, a confirmation according to which if time machines were possible, we would see "hordes of tourists from the future" (610). Moreover, Hawking indicates physical results, which come from both quantum theory and general relativity, in order to rule out travels into the past.

In the general relativity realm, the very first solution of Einstein's gravitational field equations that predicts CTCs was the van Stockum solution in the thirties.[5] However, the existence of such curves in the van Stockum space-time was only pointed out by Tipler years later.[6] The impressive Gödel[7] universe as well as the two non-intersecting cosmic strings of Gott[8] present CTCs as well. Therefore, as we can see, Einstein's theory of gravity provides *naturally* such curves, and a strong, definitive, and general argument that rules out time machines and their paradoxes is missing in physics so far.[9]

---

[4] Hawking, S. W. "Chronology protection conjecture." *Physical Review D* 46 (2): 603-611, 1992.

[5] van Stockum, W. J. "The gravitational field of a distribution of particles rotating about an axis of symmetry." P*roceedings of the Royal Society of Edinburgh* 57: 135-154, 1938. A review on CTCs in general relativity is found in Lobo, F. S. "Closed timelike curves and causality violation." In Vincent R. Frignanni (Ed.), *Classical and quantum gravity: theory, analysis and applications*. New York: Nova Science Publisher, 2012, and Earman, J., Smeenk, C., and Wüthrich, C. "Do the laws of physics forbid the operation of time machines?" *Synthese* 169 (1): 91-124, 2009. The latter discusses some time machine paradoxes and questions on the possibility of generating CTCs in accordance with laws of physics.

In this paper, I present a philosophical argument in order to deny the physical reality of CTCs or time machines. According to the argument, time machines are forbidden because individuation is an uninterrupted process. That is, following Friedrich Nietzsche and Heraclitus' notion of becoming, and Gilbert Simondon's concept of process of individuation, it is shown that individuation—when not considered as a process—leads to the belief in isolated entities or milieu-independent entities, and, consequently, provides the belief in time machines. Accordingly, human individuals considered as isolated entities will be fictional because humans are generated by processes of individuation and are immersed in collective contexts. It is worth to emphasize that the argumentation in this article is philosophical one. In order to deny time machines (focusing on travels into the past), I will not use physical and mathematical concepts. I will emphasize the concept of individuation instead of physical-mathematical concepts as indicated, for example, in Earman et al. As we will see, the novel argument presented in this work is based on the concept of process of individuation. At the end of this article, it is suggested an origin for the time machines fable: the modern *ressentiment*. And in the *amor fati* formula of Nietzsche we find the antipode to the modern *ressentiment* and time machines.

---

[6] Tipler, F. J. "Rotating cylinders and the possibility of global causality violation." *Physical Review D* 9 (8): 2203-2206, 1974.

[7] Gödel, K. "An example of a new type of cosmological solutions of Einstein's field equations of gravitation." *Reviews of Modern Physics* 21 (3): 447-450, 1949.

[8] Gott, J. R. "Closed timelike curves produced by pairs of moving cosmic strings: exact solutions." *Physical Review Letters* 66 (9): 1126-1129, 1991.

[9] Physical arguments against time machines have been pointed out as well. In Introduction, we saw Hawking's argument. But, even in physics there are others. It is shown that Gott's time machine has non-physical origin or source in Deser, S., Jackiw, R., and 't Hooft, G. "Physical Cosmic Strings Do Not Generate Closed Timelike Curves." *Physical Review Letters* 68: 267-269, 1992. On the other hand, in Pavan, A. B., Abdalla, E., Molina, C. "Stability, causality, and quasinormal modes of cosmic strings and cylinders." *Physical Review D* 81 (4): 044003, 2010, the authors show that CTCs are unstable and unable to promote travels in time. However, those studies are not general, they were made from both a specific background, or class of space-times, and a particular quantum field, the scalar field in Pavan et al. Even Hawking's criticism depends on either the non-acceptance of the weak energy condition violation or the validity of large back-reaction effects, which would prevent formation of CTCs. But today in physics, violations in energy conditions are more acceptable since the detection of cosmic acceleration, promoted by dark energy, and back-reaction effects are speculations from an incomplete quantum theory of gravity. Thus, a general and convincing physical argument against CTCs is absent in physics today.

**The Problem of Individuation**

I construct an argument against time machines from the philosophical concept of individuation, or generation of an individual. For various thinkers in the history of philosophy, individuation has an origin: a supposed principle of individuation (*principium individuationis*). The principle of individuation was a very useful concept adopted by philosophers. *The Cambridge Dictionary of Philosophy*, for example, says that the principle of individuation is "what makes something individual as opposed to universal" (737). In this sense, a specific man is different from the universal man because of the principle of individuation. Then, according to this perspective, our world is *made up* of various entities, or individuals, for the principle of individuation is present.

In Schopenhauer's philosophy, the principle of individuation promotes the world as representation. As the thing-in-itself is the will, something beyond the principle of individuation,[10] Schopenhauer claims that such a principle generates individuals from the will, or unity, which is the origin of the world, "the innermost essence, the kernel, of every particular thing and also of the whole."[11] Therefore, unity, or the will, presents itself as a myriad of objects, i.e., our physical world is the will by means of the principle of individuation.

In the same direction, for the young Nietzsche, the principle of individuation is identified to a drive (*Trieb*), which receives the name of the Greek God Apollo. But the origin of the world, such as in Schopenhauer's work, is attributed to unity, which in Nietzsche's initial philosophy is the Primordial Unity (*Ur-Eine*) (BT I). The mature Nietzsche rejected Schopenhauer's influence, and, from the maturity period, the philosopher created his own concepts, like will to power. Even the Primordial Unity was ruled out, because the world as wills to power is conceived of as plural.[12] Above all, the metaphysical principle of individuation was the attempt of describing the multiplicity in terms of a unique origin, or a unique cause, for those important thinkers. Thus, individuation and its supposed cause, the principle of individuation, are ingredients in order to justify the physical world from a metaphysical origin.

---

[10] In *The World as Will and Representation*, the principle of individuation is equivalent to space and time: "I shall call time and space the *principium individuationis*, an expression borrowed from the old scholasticism (...)" (Second Book, 23).

[11] Ibid, Second Book, 21.

[12] See Müller-Lauter, W. "Nietzsche's Teaching of Will to Power," translated by Drew E. Grin. *Journal of Nietzsche Studies*, 4/5: 37-101, 1993.

Without metaphysical speculations, individuation may be seen as a *natural process*. Instead of a metaphysical principle, individuation may be conceived of as a process, described by physical and mathematical concepts, according to the French philosopher Gilbert Simondon. Thus, to think of Simondon's individuation means to realize a process with degrees: "I intend therefore to study the *forms, modes and degrees of individuation* in order to situate accurately the individual in the wider being (…)," said Simondon.[13] According to Simondon, an individual is a process of individuation acting. An individual, in his perspective, is a metastable system that comes from another metastable system: a preindividual system. Both living beings and physical objects are systems with non-vanishing potential energy, are processes, i.e., are non-static beings:[14]

> The process of individuation must be considered primordial, for it is this process that at once brings the individual into being and determines all the distinguishing characteristics of its development, organization and modalities. Thus, the individual is to be understood as having a relative reality, occupying only a certain phase of the whole being in question — a phase that therefore carries the implication of a preceding preindividual state, and that, even after individuation, does not exist in isolation, since individuation does not exhaust in the single act of its appearance all the potentials embedded in the preindividual state. Individuation, moreover, not only brings the individual to light but also the individual-milieu dyad.

For Simondon, the process of individuation, or individuation, does not generate an isolated being. The individual-milieu dyad also appears during the process. In Simondon, we find a description of the problem of individuation as uninterrupted process. Individuals are neither static beings nor isolated entities without any relation with the environment (milieu): a full, complete, and isolated individual is something fictitious. In relation to living beings, the process of individuation acts continuously, individuating itself: "The living being resolves its problems not only by adapting itself which is to say, by modifying its relationship to its milieu (something a machine is equally able to do) — but by modifying itself through the invention of new internal structures (…)."[15] Following Simondon, Weinbaum and Veitas[16] proposed a new form to define and conceive of intelligence by means of the process of individuation. For those authors, intelligent agents emerge from a complex context and become intelligent from a process of self-organization and formation, where "individuation is a resolution of a problematic situation" (381).

---

[13] See Simondon, G. *The Genesis of the Individual*, translated by Mark Cohen and Sanford Kwinter. In Crary, J. and Kwinter, S. (eds.). Incorporations (Zone 6). New York: Zone Books, 1992, p. 311.

[14] Ibid, p. 300.

In my point of view, Simondon's process of individuation carries concepts and similar interpretations to Heraclitus and Nietzsche's philosophy of becoming. Such as Heraclitus' world view (or at least the Platonic Heraclitus[17]), we can see the importance of becoming, or process, in Simondon's philosophy. Denying any eternal substance, Simondon says that "the opposition between being and becoming can only be valid within a certain doctrine that supposes that the very model of being is a substance."[18] Such as in Nietzsche's philosophy, we may find the question about "stability" of individuals. In the mature Nietzschean philosophy, individuals are transitory configurations of wills to power.[19]

In this article, the main argument in order to deny time machines—especially travels into the past—is found in an image from Heraclitus of Ephesus, and images, or metaphors, have been useful in science as well. In Thermodynamics, for example, the volume of a perfect gas may be thought of as a set of non-interacting little balls. In general relativity, the space-time curvature may be suggested by a heavy body upon a trampoline, deforming its surface. In particular, a metaphor underlies hypothetical time travels: it is the metaphor of a time traveler as a "free particle," i.e., a time traveler playing the role of a non-interacting particle. As we will see, an isolated individual, generated by a hypothetical principle of individuation or, equivalently, a complete process of individuation is the origin for such a metaphor. Then, the metaphorical ingredient is present to think of (or to construct) *reality* in the most abstract natural science as well.

---

[15] Ibid, p. 305.

[16] Weinbaum, D. and Veitas, V. "Open ended intelligence: the individuation of intelligent agents." J*ournal of Experimental and Theoretical Articial Intelligence* 29 (2): 371-396, 2017.

[17] Cf. Kahn, C. H. *The art and thought of Heraclitus*. Cambridge: Cambridge University Press, 1979.

[18] See Simondon, G. "The position of the problem of ontogenesis," translated by Gregory Flanders. *Parrhesia* 7: 4-16, 2009, p. 6.

[19] See Müller-Lauter, W. "Nietzsche's Teaching of Will to Power," translated by Drew E. Grin. *Journal of Nietzsche Studies*, 4/5: 37-101, 1993.

Heraclitus, "the philosopher of becoming," is supposed to say, according to Plato, "that all things are in motion and nothing at rest (...)."[20] The philosopher of Ephesus compared all things "to the stream of a river" and said "that you cannot go into the same water twice."[21] Not only the river flows, but everything is in flux, even the observer who observes the flux: "In the same river, we both step and do not step, we are and we are not".[22] Thus Heraclitus denied stability for the entire world. For Nietzsche, above all, Heraclitus denied the concept of being as something static, considering it illusory and fictitious. The German philosopher and philologist, Nietzsche, was a hard critic of the philosophical language. Behind philosophical concepts, Nietzsche saw prejudgments and idiosyncrasies. Nietzsche, in the mature period of his work, criticized philosophy and its dogmas, philosophers and their bias and prejudgments. By means of a strong language criticism, the German philosopher attacked the foundations of philosophy and science. And the notions of both thing and the thing-in-itself are within his criticism. In a fragment of 1887, Nietzsche wrote (KSA 12:10[202]):

> The "thing-in-itself" absurd. If I think away all the relationships, all the "qualities", all the "activities" of a thing, then the thing does *not* remain behind: because thingness was only a "fiction added" by us, out of the needs of logic, thus for the purpose of designation, communication (...).

The German philosopher considered the concept of thing-in-itself absurd because every "thing," as language construction, depends on humans, it is related to humans. Nietzsche also criticized the notion of thing in a passage in which mathematics and the classical logic, or the principle of identity, are attacked:

> The invention of the laws of numbers was made on the basis of the error, dominant even from the earliest times, that there are identical things (but in fact nothing is identical with anything else); at least that there are things (but there is no "thing") (HH 19).

According to Nietzschean philosophy, a specific "thing" is a human creation. More specifically, it is a creation from the human body—it depends on the body structure. Kant's forms of sensibility and the understanding, which are conditions to think of and to know an object, are transferred to the human body in Nietzsche's philosophy.[23] For Nietzsche, a given thing is an interpretation of becoming. Nietzsche followed Heraclitus and adopted the same point of view of the Ephesus thinker. That is, Nietzsche's world view in his maturity is Heraclitean in some degree. The world is interpreted as becoming (*Werden*), i.e., *the nature of the world* is change, process, flux or wills to power struggling: "becoming, effecting, is only a result" of wills to power (KSA 13:14[79]). In another fragment, the philosopher wrote: "All that happens, all movement, all becoming as a determining of relations of degree and force, as a *struggle* (...)" (KSA 12:9[91]). That is, "*This world is the will to power — and nothing besides!*" (*KSA* 11:38[12]).[24] Therefore, Nietzsche assumes Heraclitus' river image or, at least, the Platonic interpretation of Heraclitus. In *Ecce homo*, the German philosopher wrote on his Dionysian philosophy and his affinity with Heraclitus' philosophy:[25]

---

[20] Cf. *Cratylus*, 402a, in Plato. *The Dialogues of Plato*, translated by B. Jowett. London: Oxford University Press, 1892.

[21] Ibid.

[22] *Heraclitus of Ephesus*, 49a, in Freeman, K. *Ancilla to the pre-Socratic philosophers: a complete translation of the fragments in Diels*. Cambridge: Harvard University Press, 1948.

> The affirmation of passing away *and destruction* that is crucial for a Dionysian philosophy, saying yes to opposition and war, *becoming* along with a radical rejection of the very concept of "being" — all these are more closely related to me than anything else people have thought so far (EH "The Birth of the Tragedy" 3).

According to his Dionysian philosophy, a given "thing," or individual, is conceived of as a "clipping" from becoming, and the concept of "being," as something stable, is only a fiction that comes from a drive that refuses the total becoming. In *Twilight of the idols* he said: "(..) Heraclitus will always be right in thinking that being is an empty fiction" (TI "Reason in philosophy" 2). Therefore, in this perspective, a totally isolated object, or individual, such as a free particle (an extreme act of individuation in physics created in order to simplify calculations because a free particle does not interact), and something stable like the Parmenidean "being" are only chimeras.

---

[23] Body in Nietzsche means mind, spirit as well: "Of all that is written I love only that which one writes with his blood. Write with blood, and you will experience that blood is spirit" (Z I "On Reading and Writing").

[24] Cf. Müller-Lauter, W. "Nietzsche's Teaching of Will to Power," translated by Drew E. Grin. *Journal of Nietzsche Studies*, 4/5: 37-101, 1993, and, for a physical point of view, Neves, J. "Cosmologia dionisíaca." *Cadernos Nietzsche* 36 (1): 267-277, 2015, and "Nietzsche for physicists." *Philosophia Scientiæ* 23 (1): 185-201, 2019.

[25] Nietzschean view on war was constructed from the Greek concept of agon. In an initial text, The Homer's contest of 1872, the young Nietzsche emphasizes the dispute as a Leitmotiv in the ancient Greek culture.

The river image, or metaphor, provides an argument against time machines. First of all, time machines, especially travels into the past, assume full processes of individuation because human travelers are conceived of as isolated, static beings, and detached "things" from becoming, from the universal flux, and from a milieu. Supposedly, the traveler through a CTC would return to another point *over the river*, interacting with another historical time and context, or another milieu as Simondon called. But in the river image, to abandon becoming and to travel into the past are impossible. The belief in some sort of complete individuation—provided by an interpretation of individuation as non-process, supposedly generated by the principle of individuation—leads to the belief in humans as something detached from becoming, or from some sort of context. The river picture, in turn, and essentially the becoming perspective reveal such a hypothetical travel as *science fiction* because the full individual—as a "free particle"[26]—and the Parmenidean concept of being are fable. The individual is not an *aeterna veritas* (eternal truth), is not apart from the universal flux, and, according to Nietzsche, this is a common error in philosophy:

> All philosophers have the common failing of starting out from man as he is now and thinking they can reach their goal through an analysis of him. They involuntarily think of "man" as an *aeterna veritas*, as something that remains constant in the midst of all flux, as a sure measure of things (HH 2).

The pictorial argument presented in this article—the becoming point of view—does not deny time dilation (or different elapsed times) described by both special relativity and general relativity.[27] Such as on the ordinary river, flux in Heraclitus' river is not invariant. In a real river, the fluid velocity depends on the position and depth. A mass or volume of water (an "individual" or "being" in this metaphor) will have different velocities if its position is close or not to the margin, or at a great depth. Different elapsed times given by Lorentz's transformations in special relativity would be assured in Heraclitus' metaphor because the universal flux is not ever the same. But CTCs or travels into the past would be ruled out.

---

[26] This is another metaphor. A "real" traveler through a CTC would experience tidal forces during a hypothetical time travel. The term "free particle" here indicates only an isolated individual from any context.

[27] In special relativity, time dilation is given by the relative motion of observers. On the other hand, in general relativity, time dilation is given by the gravitational redshift, which is generated by variations in the gravitational field and provides the GPS (Global Positioning System) technology.

The dear reader could ask me about the possibility of traveling over an ordinary or real river. By using a boat, for example, a traveler could reach any point over a real river. However, in this example, the boat does not make part of the universal flux. In this argument, the boat is considered as something firm, as a stable being—something different from becoming. But Heraclitus taught us that "all things are in motion" (including that boat), or that all things flow, and thus spoke Zarathustra in a brilliant passage in which the Ephesus philosopher and his doctrine are indicated (Z III "On Old and New Tablets" 8):

> If timbers span the water, if footbridges and railings leap over the river, then surely the one who says "Everything is in flux" has no credibility.
>
> Instead, even the dummies contradict him. "What?" say the dummies, "everything is supposed to be in flux? But the timbers and the railings are *over* the river!
>
> *Over* the river everything is firm, all the values of things, the bridges, concepts, all 'good' and 'evil' — all of this is *firm*!" —
>
> But when the hard winter comes, the beast tamer of rivers, then even the wittiest learn to mistrust, and, sure enough, then not only the dummies say: "Should everything not — *stand still*?"
>
> "Basically everything stands still" — that is a real winter doctrine, a good thing for sterile times, a good comfort for hibernators and stove huggers.
>
> "Basically everything stands still" — but *against this* preaches the thaw wind!
>
> The thaw wind, a bull that is no plowing bull — a raging bull, a destroyer that breaks ice with its wrathful horns! But ice — *breaks footbridges*!
>
> Yes my brothers, is everything not *now in flux*? Have all railings and footbridges not fallen into the water? Who could still *hang on* to "good" and "evil"?
>
> "Woe to us! Hail to us! The thaw wind is blowing!" — Preach me this, oh my brothers, in all the streets!

***Amor Fati* Contra Time Machines**

Considering modernity, Nietzsche's diagnosis is clear: the phenomenon of *ressentiment* (resentment) is present. Then, it is not difficult to identify the reason why modern man believes in time machines. Firstly, modernity emphasizes individuation, insofar as it created the image of citizens as social atoms. However, as we saw, such an image is not appropriate when we adopt the notion of process of individuation. Secondly, that man of *ressentiment* believes that it is possible to travel into the past not only in order to witness historical events but in order to correct them. In *The Birth of Tragedy*, Nietzsche describes that type of man, whose archetype is Socrates, who believes "that thought, as it follows the thread of causality, reaches down into the deepest abysses of being, and that it is capable, not simply of understanding existence, but even of *correcting* it" (BT 15). I argue that the time machine fable is, above all, created by *ressentiment*, i.e., the phenomenon of *ressentiment* is the origin for hypothetical travels into the past. Such a fable comes from the impossibility of accepting fate. On the other hand, it is by means of the *amor fati* formula that Nietzsche denies the modern *ressentiment* and (likely) would reject time machines:

> My formula for human greatness is *amor fati*: that you do not want anything to be different, no forwards, not backwards, not for all eternity. Not just to tolerate necessity, still less to conceal it — all idealism is hypocrisy towards necessity — , but to *love* it... (EH "Why I am so Clever" 10).

Han-Pile emphasizes that the Nietzschean formula assumes "a transformation, not of the past, but of *ourselves*" (242). Then we may conceive of *amor fati* as an alternative to time machines (and a criticism), i.e., the *amor fati* formula of Nietzsche "represents a human, heteronomous alternative to willing backwards (…)," according to Han-Pile (243).

By associating time machines—in particular travels into the past—with the man of *ressentiment*, it is possible to advocate, as Nietzsche indicated in *On the Genealogy of Morality*, the lineage of modern science, which points toward the priest (the archetype or personification of *ressentiment*, according to Reginster[28]), a representative of the ascetic ideal. For priests, the ascetic ideal means "the actual priestly faith, their best instrument of power and also the "ultimate" sanction of their power" (GM III 1). Thus science, according to Nietzsche, "is not the opposite of the ascetic ideal but rather the latter's own *most recent and noble manifestation*" (GM III 23), and the time machine fable suggests such a relation.

---

[28] Cf. Reginster, B. "Nietzsche on ressentiment and valuation." *Philosophy and Phenomenological Research* 57 (2): 281-305, 1997, p. 289.

**Final Remarks**

Closed time-like curves (CTCs), or time machines, are objects within Einstein's theory of general relativity. The van Stockum space-time and the Gödel cosmological model are examples of solutions of Einstein's field equations that possess CTCs. Researches have proposed physical and philosophical arguments in order to exclude time machines as physical realities and its inherent paradoxes. However, a strong, general, and persuasive argument that rules out time travels, especially travels into the past, is missing in physics and philosophy so far. Arguments like Hawking's chronology protection conjecture show that ingredients beyond general relativity are necessary in order to reject time machines. Therefore, this paper presents an ontological argument against time machines that comes from the philosophy of becoming. Individuation thought of as a non-process, supposedly generated by a metaphysical principle of individuation, motivates the belief in human beings who would travel into the past. In an exaggerated degree, individuation as non-process gives rise to the belief in human beings conceived of as isolated beings, as something separated from their contexts and milieus. However, an individual emerges from a society, culture with values and language, and from a specific historical time. The belief in the complete individuation ignores such a condition, and one dreams with humans as "free particles" traveling into the past. Then, by using Simondon's process of individuation, Heraclitus' river image, in which "everything is in flux," and Nietzsche's "radical rejection of the very concept of 'being'," time machines appear as subject of science fiction, and CTCs become *non-physical* objects of Einstein's theory of gravity. Lastly, I proposed a psychological origin for the time machine fable. The modern *ressentiment* gives rise to the time travel fable, in which a hypothetical and optimistic time traveler would correct all historical events. Then, the *amor fati* formula of Nietzsche is a response to the resentful time traveler.

## Works Cited

Abbott, B. P. et al. (LIGO Scientific Collaboration and Virgo Collaboration). "Observation of gravitational waves from a binary black hole merger." *Physical Review Letters* 116: 061102, 2016.

Audi, R. (ed.). *The Cambridge dictionary of philosophy*. New York: Cambridge University Press, 1999.

Brandenberger, R., and Peter, P. "Bouncing cosmologies: progress and problems." *Foundations of Physics, 47* (6): 797-850, 2017.

Deser, S., Jackiw, R., and 't Hooft, G. "Physical Cosmic Strings Do Not Generate Closed Timelike Curves." *Physical Review Letters* 68: 267-269, 1992.

Earman, J., Smeenk, C., and Wüthrich, C. "Do the laws of physics forbid the operation of time machines?" *Synthese* 169 (1): 91-124, 2009.

Freeman, K. *Ancilla to the pre-Socratic philosophers: a complete translation of the fragments in Diels.* Cambridge: Harvard University Press, 1948.

Gödel, K. "An example of a new type of cosmological solutions of Einstein's field equations of gravitation." *Reviews of Modern Physics* 21 (3): 447-450, 1949.

Gott, J. R. "Closed timelike curves produced by pairs of moving cosmic strings: exact solutions." *Physical Review Letters* 66 (9): 1126-1129, 1991.

Han-Pile, B. "Nietzsche and amor fati." *European Journal of Philosophy* 19 (2): 224-261, 2009.

Hawking, S. W. "Chronology protection conjecture." *Physical Review D* 46 (2): 603-611, 1992.

Kahn, C. H. *The art and thought of Heraclitus.* Cambridge: Cambridge University Press, 1979.

Lobo, F. S. "Closed timelike curves and causality violation." In Vincent R. Frignanni (Ed.), *Classical and quantum gravity: theory, analysis and applications.* New York: Nova Science Publisher, 2012.

Müller-Lauter, W. "Nietzsche's Teaching of Will to Power," translated by Drew E. Grin. *Journal of Nietzsche Studies*, 4/5: 37-101, 1993.

Neves, J. "O eterno retorno hoje." *Cadernos Nietzsche* 32: 283-296, 2013.

— "Cosmologia dionisíaca." *Cadernos Nietzsche* 36 (1): 267-277, 2015.

— "Bouncing cosmology inspired by regular black holes." *General Relativity and Gravitation* 49 (9): 124, 2017.

— "Nietzsche for physicists." *Philosophia Scientiæ* 23 (1): 185-201, 2019.

Nietzsche, F. *Samtliche Werke. Kritische Studienausgabe.* Berlin-New York: Walter de Gruyter, 1978.

— *Writings from the late notebooks*, translated by Kate Sturge. Cambridge: Cambridge University Press, 2003.

— *Human, all too human*, translated by R. J. Hollingdale. Cambridge: Cambridge University Press, 2005.

—*The anti-Christ, Ecce homo and Twilight of the Idols*, translated by Judith Norman. Cambridge: Cambridge University Press, 2005.

—*Thus spoke Zarathustra*, translated by Adrian del Caro. Cambridge: Cambridge University Press, 2006.

—*The Birth of Tragedy and other writings*, translated by Ronald Speirs. Cambridge: Cambridge University Press, 2007.

—*On the Genealogy of Morality*, Cambridge: Cambridge University Press, translated by Carol Diethe, 2007.

Novello, M. and Perez Bergliaffa S. E. "Bouncing cosmologies." *Physics Reports* 463 (4): 127-213, 2008.

Pavan, A. B., Abdalla, E., Molina, C. "Stability, causality, and quasinormal modes of cosmic strings and cylinders." *Physical Review D* 81 (4): 044003, 2010.

Plato. *The Dialogues of Plato*, translated by B. Jowett. London: Oxford University Press, 1892.

Reginster, B. "Nietzsche on ressentiment and valuation." *Philosophy and Phenomenological Research* 57 (2): 281-305, 1997.

Schopenhauer, A. *The world as will and representation*, translated by E. F. J. Payne. New York: Dover Publications, Inc, 1969.

Simondon, G. *The Genesis of the Individual*, translated by Mark Cohen and Sanford Kwinter. In Crary, J. and Kwinter, S. (eds.). Incorporations (Zone 6). New York: Zone Books, 1992.

— "The position of the problem of ontogenesis," translated by Gregory Flanders. *Parrhesia* 7: 4-16, 2009.

Tipler, F. J. "Rotating cylinders and the possibility of global causality violation." *Physical Review D* 9 (8): 2203-2206, 1974.

van Stockum, W. J. "The gravitational field of a distribution of particles rotating about an axis of symmetry." P*roceedings of the Royal Society of Edinburgh* 57: 135-154, 1938.

Weinbaum, D. and Veitas, V. "Open ended intelligence: the individuation of intelligent agents." J*ournal of Experimental and Theoretical Artificial Intelligence* 29 (2): 371-396, 2017.